\documentclass[twocolumn,nofootinbib,prd,aps,showpacs,preprintnumbers,
tightenlines]{revtex4}
\usepackage{epsfig} 
\begin{document}
\title{Neutral pion decay in dense skyrmion matter}
\author{Alexander C. Kalloniatis$^a$}
\email{akalloni@physics.adelaide.edu.au}
\author{Byung-Yoon Park$^{a,b}$}
\email{bypark@cnu.ac.kr}
\affiliation{$^a$ CSSM, University of
Adelaide, Adelaide 5005, Australia
 \\
$^b$ Department of Physics,Chungnam National University, 
Daejon 305-764, Korea }
\date{\today}
\preprint{ADP-04-25/T607}
\begin{abstract}
We study the density dependence of the decay 
$\pi^0\rightarrow \gamma \gamma$ using  
the Skyrme Lagrangian to describe simultaneously 
both the matter background and mesonic fluctuations. 
The classical ground state configuration has different 
chiral properties depending on the skyrmion density, 
which is reflected in the physical properties of pion 
fluctuating on top of the classical background.
This leads to large suppression at high density
of both photo-production from the neutral pion 
and the reverse process.
The effective charges of $\pi^{\pm}$ are also 
discussed in the same framework.
\end{abstract}

\maketitle

\section{Introduction}
At high temperature and/or density, the properties of hadrons are 
expected to
change dramatically and understanding these changes  
under extreme conditions is important not only in
nuclear and particle physics but also in many other related fields
such as astrophysics. Data from high-energy heavy ion
colliders, astronomical observations on compact stars and some
theoretical considerations suggest that the phase diagram of
hadronic matter is far richer than the simple confinement/deconfinement
picture seen in finite temperature lattice QCD simulations 
and include interesting QCD phases such as color superconductivity. 
Moreover effective theories
can be derived for these extreme conditions, using macroscopic
degrees of freedom, by matching them to QCD at a scale close to
the chiral scale $\Lambda_\chi \sim 4\pi f_\pi \sim 1$ GeV.

Chiral symmetry, which under
normal conditions is spontaneously broken, is believed to be
restored under such extreme conditions by virtue of its seeming to
go hand-in-hand with confinement. The value of the quark
condensate $\langle \bar{q} q \rangle$ of QCD is an order
parameter of this symmetry and is expected to decrease as the
temperature and/or density of hadronic matter are increased. Since
the pion is the Goldstone boson associated with spontaneously broken
chiral symmetry, the various patterns in which the symmetry is
realised in QCD will be directly reflected in the in-medium properties of
the pion, such as its mass $m_\pi$ and decay decay constant $f_\pi$. 

These quantities have been the subject of previous 
studies\cite{LPMRV03,LPRV03,LPRV04,PRV04} in the
formalism adopted here.
In those works, the Skyrme picture is used
to describe both pions and dense baryonic
matter. The basic strategy of the approach begins with the Skyrme  
conjecture that a soliton (skyrmion) 
of the meson Lagrangian can be taken as a baryon, 
so that dense baryonic matter can be approximated as 
a system of infinitely many skyrmions. 
The pion is then incorporated as a fluctuation over  
this dense skyrmion matter. 
The chiral properties of the classical background solution 
is directly reflected in the pion fluctuation properties, 
which may be interpreted as in-medium modifications.  
If we accept that the Skyrme model can be applied up to 
some density, its unique feature of a unified meson-baryon 
description provides an interesting framework to investigate 
nonperturbatively the meson properties in dense baryonic matter.
That is, we do not have to assume any density dependence of 
the in-medium parameters. We work with a single model Lagrangian 
whose parameters are fixed for mesons in free space. 
Only the classical ground state describing the dense skyrmion matter 
becomes highly density dependent and thus naturally in turn
so do the fluctuating mesons on top of this dense background.
There are however a few drawbacks in the approach. Firstly, the 
lowest energy configurations for skyrmionic  
matter are only available thus far for a 
crystal structure~\cite{SkyrmionCrystal};
we cannot yet describe the liquid structure of normal nuclear matter
nor its behaviour at high temperature. 
Next, there are a few undetermined parameters in the model. 
They limit us to a qualitative understanding of the related physics. 
Leaving these drawbacks to further improvements, 
in this paper we apply the approach to the anomalous decay of the 
neutral pion into two photons, $\pi^0\rightarrow \gamma \gamma$
at finite density. 
Though this electromagnetic process may not be regarded 
as so important in the fireball phase of relativistic heavy ion 
collisions,
it is potentially relevant, for example,
through $\gamma\gamma \rightarrow \pi^0 \rightarrow \nu\bar{\nu}$,  
to astrophysical phenomena such as core collapse supernovae and neutron 
stars \cite{Astrophysics,ANV03}. 

The dependence of the process $\pi^0 \rightarrow \gamma\gamma$ 
on temperature has been investigated by various 
authors \cite{FiniteT,Pi96,PTT97,AEDGN94}. 
It is well understood that though the anomaly, which
drives the decay, is temperature independent, nonetheless
the amplitude does depend on $T$ through the phase space 
or through the modification of the thermal quark propagators 
in evaluating explicitly the triangle diagram.
In particular, \cite{PTT97,AEDGN94} consider temperature
effects in the WZW effective Lagrangian which incorporates
the anomaly. The coefficient of the anomalous term 
corresponding to $\pi^0\rightarrow \gamma\gamma$ can pick up 
temperature dependence from loops \cite{Pi96}, which is close in
spirit to what will emerge in our study of density dependence below
though for us the coefficient modifications come
not from loops but at tree-level in the background field approach.  

Several works have studied also the density dependence
of the process, with quite different results emerging
\cite{GNAE93,CRK03,Caldas,CFL}. In \cite{GNAE93} the decay
is computed from the triangle diagram but keeping
pion quantities such as mass and coupling fixed
at their {\it in vacuo} values with the result
that the decay width increases with density.
In \cite{CRK03} a diagrammatic approach in the three-flavour
NJL model is used with the result that the decay width
decreases with density despite a corresponding increase
in the pion mass. 
In \cite{Caldas} the neutral pion decay width was 
computed by assuming its free space form from the anomaly
and inputting two, quite different, model  
(such as Nambu--Jona-Lasinio) scenarios 
for the behaviour of the pion mass and decay constant at the 
phase transition point in temperature or density with
correspondingly different results, increasing and
decreasing, for the decay width dependence on density. 

In distinction to this, by using the
Skyrme Lagrangian we obtain all in-medium 
quantities such as pion mass and decay constant 
from the same Lagrangian from which we compute the
decay width. We stress that though our approach
uses some of the machinery of the Skyrme approach,
the most important aspect of the calculation is
that the density dependence is obtained from the
{\it same effective chiral Lagrangian} as that for free space
using the fluctuation formalism. Other details
of the Skyrme model do not play a significant role.  

In \cite{CFL}, a similar process
$\tilde{\pi}^0 \rightarrow \tilde{\gamma}\tilde{\gamma}$
was analysed for the generalized pion fluctuations and the 
generalized photons in the color-flavour-locked phase using the
corresponding Wess-Zumino-Witten term \cite{Hong} for that phase.
It was shown that the decay of the generalised pion is
constrained by geometry and vanishes at large density.
The color-flavour-locked phase can be taken as the high density 
limit of our approach, in the region
where a hadronic description is no longer valid. 

The paper is structured as follows: the next section
reviews the Skyrme Lagrangian and the properties of the 
dense skyrmion matter. In section III we consider
meson fluctuations on this matter background
and extract the in-medium modifications to the pion
observables. There is a brief statement of conclusions
in section IV.

\section{Model Lagrangian and dense skyrmion matter}
We begin with a modified Skyrme model Lagrangian 
\cite{ScaledLagrangian}, which
also incorporates the scale anomaly of QCD in terms
of a scalar dilaton field:
\begin{eqnarray}
{\cal L} &=& 
\frac{f_\pi^2}{4}
  \left( \frac{\chi}{f_\chi} \right)^2
  \mbox{Tr} (\partial_\mu U^\dagger \partial^\mu U) 
 + {\cal L}_{\mbox{\scriptsize sk}}
\nonumber\\
&&+\frac{f_\pi^2 m_\pi^2}{4} 
  \left( \frac{\chi}{f_\chi} \right)^3
  \mbox{Tr} (U^\dagger + U - 2) 
 + {\cal L}_{\mbox{\scriptsize WZW}}  
\label{L0}\\
&& 
+ \frac{1}{2} \partial_\mu \chi \partial^\mu \chi
- \frac{m_\chi^2 f_\chi^2}{4} \left(
({\chi}/{f_\chi})^4(\mbox{ln}(\chi/f_\chi)-\textstyle\frac14)
 + \frac14 \right)
\nonumber 
\end{eqnarray}
where $U=\exp(i\vec{\tau}\cdot\vec{\pi}/f_\pi) \in SU(2)$ and
$\chi$ is the scalar dilaton field. 
The parameters in Eq.~(\ref{L0}) correspond to physical 
properties of the corresponding mesons: 
$m_\pi$ and $f_\pi$ ($m_\chi$ and $f_\chi$) 
are the pion (respectively, dilaton) mass and decay constant in free space. 
The Skyrme term ${\cal L}_{\mbox{\scriptsize sk}}$ 
is the higher derivative term introduced into the Lagrangian 
to stabilise the soliton solution of the Lagrangian, namely
\begin{equation}
{\cal L}_{\mbox{\scriptsize sk}} 
= \frac{1}{32e^2} \mbox{Tr}
\left( [ L_\mu, L_\nu]^2 \right),
\end{equation}
where $L_\mu \equiv (\partial_\mu U) U^\dagger$
($R_\mu \equiv U^\dagger (\partial_\mu U)$).
Finally, the Wess-Zumino-Witten term 
${\cal L}_{\mbox{\scriptsize WZW}}$ is necessary 
to break the symmetry of Eq.(\ref{L0}) under $U\rightarrow U^{\dagger}$  
which is not a genuine symmetry of QCD. 
The corresponding action can be written locally as \cite{Witten83}
\begin{equation}
  S_{\mbox{\scriptsize WZW}} = 
     -\frac{iN_c}{240\pi^2} 
     \int \varepsilon^{\mu\nu\lambda\rho\sigma} d^5 x 
  \mbox{Tr} \left( L_\mu \cdots L_\sigma \right)
\label{WZW0}\end{equation}
in a five-dimensional space whose boundary is ordinary 
space and time. For $U\in SU(2)$, namely for two flavors,
$S_{\mbox{\scriptsize WZW}}$ trivially vanishes
(for three flavors this gives the hypothesised
process $KK\rightarrow \pi \pi \pi$). 
This will change when we couple to photons, as discussed below.  

The model Lagrangian has a few parameters which are fixed by 
the meson dynamics in baryon-free space. For the pions, we can 
fix the associated parameters to the empirical values 
as $f_\pi = 93$ MeV and $m_\pi=138$ MeV. 
On the other hand, for the dilaton field, there are no available 
well-established empirical values for the mass and decay constant 
(or equivalently the vacuum expectation value). There have been 
a number of theoretical discussions on the field itself \cite{Vento03}.
We use the values phenomenologically determined in 
nuclear matter studies~\cite{dilaton}.
There is another unknown parameter which has not been determined 
by any directly associated experiment, the Skyrme parameter $e$, 
for which we use the conventionally used value~\cite{JM83,ANW83}.
Due to the ambiguities in these model parameters our study should 
be taken as having only qualitative relevance. 

The Lagrangian is invariant under global charge rotations, 
$U \rightarrow U + i\varepsilon[Q,U]$, where $\varepsilon$ 
is a constant and $Q=\mbox{diag}(\frac23,-\frac13)$ is the 
charge matrix for light quarks. The coupling of the 
pions and the photons can be incorporated by promoting this 
to a local symmetry, $\delta U \rightarrow \varepsilon(x) [Q,U].$
This can be done by replacing the 
derivatives acting on the pion fields (not the neutral 
dilaton field) 
by covariant derivatives, 
$\partial_\mu \rightarrow D_\mu = \partial_\mu + ieA_\mu$
where the photon field $A_\mu(x)$ 
transforms as 
$A_\mu \rightarrow A_\mu - (1/e) \partial_\mu \varepsilon$
and $e$ is the charge of the proton. 
For example, the current algebra term for the pions is rewritten via 
\begin{equation}
  \begin{array}{l}
     \displaystyle 
     \frac{f_\pi^2}{4}
     \left( \frac{\chi}{f_\chi} \right)^2
     \mbox{Tr} (\partial_\mu U^\dagger \partial^\mu U) 
  \\
     \hskip 5em \Rightarrow \displaystyle 
     \frac{f_\pi^2}{4} 
     \left( \frac{\chi}{f_\chi} \right)^2
     \mbox{Tr} (D_\mu U^\dagger D^\mu U).
  \end{array}
  \label{Lkin}
\end{equation}
On the other hand, as is well-known, minimal substitution
does not work in gauging the 
Wess-Zumino action. The so-called trial and error Noether 
method \cite{Witten83} gives the gauged Wess-Zumino-Witten action 
\begin{equation}
  \begin{array}{l}
     \displaystyle 
     \tilde{\Gamma}_{\mbox{\scriptsize WZW}}(U,A_\mu) 
     = 
     {\Gamma}_{\mbox{\scriptsize WZW}}(U) 
     - e \int d^4 x A^\mu J_\mu^{\mbox{\scriptsize an}} 
   \\
     \hskip 4em \displaystyle
     + \frac{ie^2}{24\pi^2} \int d^4 x 
     \varepsilon^{\mu\nu\alpha\beta} (\partial_\mu A_\nu) A_\alpha 
   \\
     \hskip 6em \displaystyle
     \times \mbox{Tr} 
     \left[ 
        Q^2 L_\beta + Q^2 R_\beta 
        + QUQU^\dagger L_\beta
     \right],
   \end{array} 
\end{equation}
where the anomalous part of the electromagnetic current of pions 
is defined as 
\begin{equation}
\begin{array}{l}
\displaystyle
  J^{\mbox{\scriptsize an}}_\mu = \frac{1}{48\pi^2} 
    \varepsilon_{\mu\nu\alpha\beta} \int d^3 x 
\\
\hskip 5em \displaystyle 
\mbox{Tr} 
  \left[Q (L_\nu L_\alpha L_\beta 
         + R_\nu R_\alpha R_\beta) \right] 
\end{array} 
\label{Jan}\end{equation}
The normal part of the current coming from the current 
algebra term (\ref{Lkin}) is 
\begin{equation}
  J^{\mbox{\scriptsize n}}_\mu 
    = i \frac{f_\pi^2}{4} 
        \left(\frac{\chi}{f_\chi^{}}\right)^2
        \mbox{Tr}\left[ Q (R_\mu - L_\mu) \right] 
  \label{Jn}
\end{equation}
and is conserved by the classical equations of motion of the fields.
 
For meson dynamics in baryon free space, 
the vacuum solutions for the pions and scalar field are  
\begin{equation}
  U_{\mbox{\scriptsize vac}}^{}=1, 
     \hskip 3em 
  \chi_{\mbox{\scriptsize vac}}^{}=f_\chi.
\end{equation}
The fluctuations on top of this free-space vacuum are then incorporated via 
\begin{equation}
  U=U_\pi=\exp(i\vec{\tau}\cdot\vec{\pi}/f_\pi),
     \hskip 1em \mbox{ and } \hskip 1em
  \chi = f_\chi + \tilde{\chi}.
  \label{Fluct0}
\end{equation}
To lowest non-trivial order we obtain  
\begin{equation}
  \begin{array}{rcl}
    {\cal L}&=& 
    \frac12 \partial_\mu \pi_a \partial^\mu \pi_a 
     - \frac12 m_\pi^2 \pi_a  \pi_a 
   \\
     &&
     + \frac12 \partial_\mu \tilde{\chi} 
               \partial^\mu \tilde{\chi}
     - \frac12 m_\chi^2  \tilde{\chi}^2
  \\
    &&
    + \frac{i}{2} e A^\mu 
        (\partial_\mu \pi^+ \pi^- 
         - \partial_\mu \pi^- \pi^+)
  \\
    && 
    \displaystyle 
    + \frac{N_c e^2}{96\pi^2 f_\pi} \pi^0
     \varepsilon^{\mu\nu\alpha\beta} 
     F_{\mu\nu} F_{\alpha\beta}
  \\  
    &&
    \displaystyle
   - \frac{ieN_c}{12\pi^2 f_\pi^3} 
     \varepsilon^{\mu\nu\alpha\beta}
     A_\mu \partial_\nu \pi^+ 
           \partial_\alpha \pi^- 
           \partial_\beta \pi^0
  \\ 
    && + \cdots.
  \end{array}
  \label{L1}
\end{equation}
Thus we generate the usual pion-photon interactions. 
Note that the normal electromagnetic current of
pions leads to the isovector current associated with
two charged pions. Correspondingly, the anomalous current
gives the isoscalar current for the neutral and charged pions. 
The two terms with $N_c$ in Eq.(\ref{L1}) come from the 
gauged Wess-Zumino-Witten term and describe the anomalous  
$\pi^0 \rightarrow \gamma\gamma$ decays and 
$\gamma \rightarrow \pi^+ \pi^- \pi^0$.  
At tree level, it leads to the standard
$\pi^0 \rightarrow \gamma\gamma$ decay width {\it in vacuo}  
\begin{equation}
\Gamma_{\pi^0 \gamma\gamma} 
= \frac{m_{\pi}^3}{64\pi} 
\left( \frac{N_c e^2}{12\pi^2 f_\pi}\right)^2.
\label{anomaly}
\end{equation}

On the other hand, 
the nonlinearity of the Lagrangian supports soliton solutions 
(skyrmions) carrying nontrivial topological winding numbers. 
With Skyrme's conjecture interpreting the winding 
number as baryon number we may simulate dense baryonic 
matter by using the meson Lagrangian as a system made of many skyrmions. 
Let the lowest energy configuration for a given baryon 
number density be 
$U_0^{}(\vec{r})\equiv n_0 + i\vec{\tau}\cdot\vec{n}$
and $\chi_0^{}(\vec{r})$.
The Classical lowest energy state of the 
multi-skyrmion system is a crystal and there has been intensive 
work in the late 80's on a model containing only pions 
\cite{SkyrmionCrystal}. 
Of course, such a crystal structure is
unrealistic for a multi-baryon system and a randomising
of the multi-skyrmion arrangement is an open problem.
There have been a few pioneering works attempting to incorporate statistical 
fluctuations (for example, see \cite{Walet02}), 
which can be adopted in our approach in principle. 
However, none of the properties we consider below will
depend explicitly on the specific periodic properties of the crystal
structure.

Two well-separated skyrmions are in a lowest energy 
configuration when relatively rotated in isospin space
about an axis perpendicular to the line joining 
their centres. We thus consider as the lowest energy state 
of skyrmionic matter at relatively low density  
that for a face centered cubic (FCC) crystal where well localised 
single skyrmions are arranged on each lattice site such that 
the twelve nearest skyrmions are each oriented corresponding to
the above lowest energy configuration for a skyrmion pair. 
At higher density, skyrmion tails start 
overlapping each other and the system undergoes a phase transition 
to a more symmetric configuration, the so called the ``half-skyrmion'' 
cubic crystal. There, one half of the baryon number 
carried by the single skyrmion is concentrated at an original FCC site
where $U_0=-1$ while the other is concentrated on the links where $U_0=+1$. 
Such a system has an additional symmetry with respect to 
$U_0 \rightarrow -U_0$, which results in the vanishing of 
$\langle U_0 \rangle$.
This is often interpreted as a restoration of the chiral symmetry
in the literature \cite{SkyrmionCrystal}.
More precisely though, it is only the average value of $U$  
over space that vanishes while the chiral circle still has a 
fixed radius $f_\pi$. 
We call this a ``pseudogap'' phase to distinguish it 
from the genuine chiral symmetry restored phase for which the 
chiral circle shrinks to a point at $U=0$.  

The dilaton field, when introduced in the Lagrangian Eq.(\ref{L0}) 
to restore scale symmetry, effectively plays the  
role of a ``radial'' field for $U$ and the restriction on 
the chiral radius is then relaxed. 
The pseudogap phase still remains as a transient 
process unless the dilaton mass is sufficiently 
small \cite{LPRV04}.
Shown in Fig.1 is a typical numerical result for the 
average value of $n_0$
and $\chi_0^{}/f_\chi$ in the lowest energy 
field configuration for a given baryon number density.  
In the figure, the baryon number density is given in units of 
normal nuclear matter density $\rho_0$. 
It should be taken only as a rough guide 
since density dependence scales strongly with    
the model parameters, especially on the ``Skyrme parameter" $e$ 
and the scalar mass $m_{\chi}$ as $\rho_p \propto e^3$ and 
$\rho_c \propto m_\chi^3$. 
One can see nevertheless that as the density increases the average value 
 $\langle n_0\rangle$ drops quickly and almost reaches zero around 
the density $\rho_p$,  where the system enters the 
pseudogap phase in the half-skyrmion configuration. 
Compared to the chiral limit $m_\pi=0$ where 
$\langle n_0\rangle$ vanishes exactly at $\rho=\rho_p$ 
(as illustrated in Fig.1 by the dashed line),  
the transition happens rather smoothly and thus
can be called an approximate or {\it quasi-} pseudogap phase. 
As we increase the density further, the system remains in this 
approximate pseudogap phase for some range of $\rho$ but the average value 
of $\chi/f_\chi$ continues slowly to decrease. At density $\rho_c$, 
the $\langle \chi/f_\chi\rangle \neq 0$ phase and 
the $\langle \chi/f_\chi\rangle = 0$ phase have the same 
energy. Then, at density higher than $\rho_c$, 
the latter comes to have lower 
energy and finally chiral symmetry is restored. 
\begin{figure}
\begin{center}
\includegraphics[width=5cm,height=7cm,angle=270]{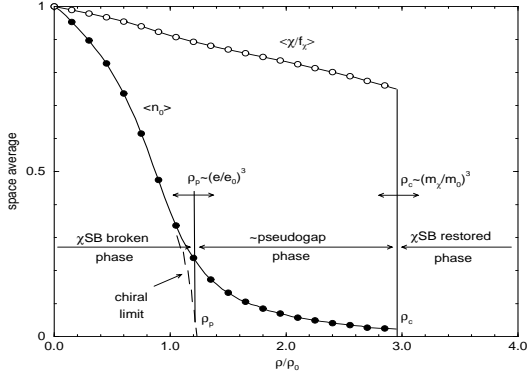}
\end{center}
\vskip -4ex
\caption{Average values of $\sigma$ and $\chi/f_\chi$
of the lowest energy crystal configuration at a given baryon number
density.}
\end{figure} 
To conclude this discussion, the dilaton in this approach
is important for setting up a consistent mechanism of chiral
symmetry restoration within the Skyrme approach. 
In the mean field approach we will use in the following 
the dilaton will not play any further role beyond this enabling a correct 
framework for change of symmetry properties approaching
the phase transition. 

\section{Fluctuations on top of the dense skyrmion matter}

We now introduce mesonic fluctuations on the dense baryonic medium
just described via 
\begin{equation}
  U = \sqrt{U_\pi} U_0 \sqrt{U_\pi}, 
    \hskip 3em 
  \chi = \chi_0 + \tilde{\chi}
\end{equation}
in Eq.(\ref{L0}). Expanding in $U_{\pi}$ we thus obtain the 
Lagrangian for the {\em in medium} fields  
\begin{equation}
  \begin{array}{rcl}
    {\cal L}
    &=&
    \textstyle 
      \frac12 G_{ab}(\vec{r}) \partial_\mu \pi_a \partial^\mu \pi_b
     -\frac12 M^2_\pi(\vec{r}) \pi^2
  \\
    && \textstyle
       +\frac12 \partial_\mu \tilde{\chi} \partial^\mu \tilde{\chi}
       -\frac12 M^2_\chi(\vec{r}) \tilde{\chi}^2 
  \\ &&
    + \frac{i}{2} e C(\vec{r}) A^\mu (\partial_\mu \pi^+ \pi^- 
                  - \partial_\mu \pi^- \pi^+)
  \\ &&
    + \displaystyle \frac{N_c e^2}{96\pi^2 f_\pi} D(\vec{r}) \pi^0
     \varepsilon^{\mu\nu\alpha\beta} F_{\mu\nu} F_{\alpha\beta}
  \\ && 
    \displaystyle
    - \frac{ieN_c}{12\pi^2 f_\pi^3} F(\vec{r}) 
      \varepsilon^{\mu\nu\alpha\beta}
      A_\mu \partial_\nu \pi^+ 
      \partial_\alpha \pi^- \partial_\beta \pi^0
  \\ && 
 + \cdots.
  \end{array}
\label{L2}
\end{equation}
where we have written only the terms corresponding to those 
in Eq.(\ref{L1}) to emphasise the in-medium modifications
of the various contributions. 
Note that this does not alter the anomaly structure 
of the Lagrangian, which is fully respected by the Lagrangian Eq.(\ref{L0}).
Only the cofficients of the terms corresponding to each process 
receive {\em local effective} 
corrections from the background matter through the potentials: 
\begin{eqnarray*}
  && G_{ab}(\vec{r}) 
     = ({\chi_0}/{f_\chi})^2 (n_0^2 \delta_{ab} + n_a n_b ), 
  \\
  && M_\pi^2(\vec{r}) = m_\pi^2 ({\chi_0}/{f_\chi})^3 n_0,
  \\
  && \begin{array}{l}
        M^2_\chi(\vec{r}) = 
          \displaystyle
          m_\chi^2 ({\chi_0}/{f_\chi})^2 
          \left[3 \ln(\chi_0/f_\chi) + 1 \right] 
        \\
         \hskip 4em
         + ({f_\pi}/{f_\chi})^2 
          [(\partial_i n_\alpha)^2
        \\
         \hskip 8em 
            + 6 m_\pi^2 (1-n_0) ({\chi_0}/{f_\chi}) 
           ],
      \end{array}
   \\
    &&  C(\vec{r}) = ({\chi_0}/{f_\chi})^2 n_0^2,
   \\
    &&  D(\vec{r}) = n_0^2 + n_3^2,
   \\
    &&  F(\vec{r}) = n_0^2.
\label{potentials} 
\end{eqnarray*}
In a mean field treatment of these potentials, we 
may take their spatial averages so that they reduce to  
constants, namely 
$\langle G_{ab}(\vec{r}) \rangle \equiv G \delta_{ab}$,
$\langle M^2_{\pi,\chi}(\vec{r}) \rangle \equiv M^2_{\pi,\chi}$,
$\langle C(\vec{r}),D(\vec{r}),F(\vec{r}) \rangle \equiv C, D, F$.
Then the Lagrangian can be rewritten simply as 
\begin{equation}
  \begin{array}{rcl}
     {\cal L} &=& 
     \textstyle
     \frac12 \partial_\mu \pi^*_a \partial^\mu \pi^*_a 
     - \frac12 m_\pi^{*2} \pi^*_a  \pi^*_a
     \\ &&
     \textstyle 
     + \frac12 \partial_\mu \tilde{\chi} \partial^\mu \tilde{\chi}
     - \frac12 m^{*2}_\chi  \tilde{\chi}^2
     \\ &&
     \displaystyle
     + \frac{i}{2} e^*_\pi A^\mu (\partial_\mu \pi^{*+} \pi^{*-} 
                  - \partial_\mu \pi^{*-} \pi^{*+}) 
     \\ &&
     \displaystyle 
     + \frac{(N_c e^{2})_{\pi\gamma^2}^*}{96\pi^2 f^*_\pi} \pi^{*0}
       \varepsilon^{\mu\nu\alpha\beta} F_{\mu\nu} F_{\alpha\beta}
     \\ &&
     \displaystyle
     - \frac{i(eN_c)_{\gamma\pi^3}^*}{12\pi^2 f_\pi^{*3}}  
       \varepsilon^{\mu\nu\alpha\beta}
       A_\mu \partial_\nu \pi^{*+} 
         \partial_\alpha \pi^{*-} 
         \partial_\beta \pi^{*0}
     \\ && 
     \cdots
  \end{array}
  \label{L*}
\end{equation}
where we have carried out a wavefunction renormalisation 
$\pi^*_a=\sqrt{\langle G\rangle} \pi_a$ for the pion fields.  
{\it All} the physical parameters are evidently modified by the  
medium. Their density dependence is given by the following relations 
\begin{eqnarray*}
   f^*_\pi/f_\pi &=& \sqrt{G}, 
   \\
   m_\pi^*/m_\pi &=& \sqrt{M^2_\pi/G}, 
   \\
   e_\pi^*/e &=& (C/G) ,
   \\
   (eN_c)^*_{\gamma\pi^3}/(eN_c) &=& F ,
   \\
   (e^2 N_c)^*_{\pi\gamma^2}/(e^2 N_c) &=& D.
\end{eqnarray*}
We now see that three different kinds of 
effective electric charge of pions
appear depending on the different electromagnetic processes involved: 
that of charged pions (denoted by $e^*_\pi$), 
that appearing in the vertices for  
$\pi^0 \rightarrow \gamma\gamma$ and thirdly that for the process  
$\gamma\rightarrow \pi^+\pi^0\pi^-$. 
All three charges were simply a unit of electric charge in 
the Lagrangian Eq.(\ref{L1}) for the pions in baryon free space. 
However, as illustrated in Fig.2, all three charges originate from 
the electric charge of the quarks and thus how each term gets   
the charge factor $e$ in its effective vertex is based on 
completely different detailed dynamics of the quarks. 
The baryonic matter influences each process in a different way so that 
each effective charge develops its own density dependence.

Above all, the ``effective'' electic charge of the charged pions 
illustrated as Fig.2(a) comes from the effect 
of the surrounding medium just as in the case of electrons in condensed matter.
As for the other processes, we also have to take into account 
not only the modification of the electric charge of the quarks 
but also the changes in the number of colors $N_c^*$ involved 
in the triangular or square anomaly diagram of Figs.3(b,c). 
The latter is expected to take place through the modifications 
of the quark propagator and the quark vacuum.  
As explicitly shown in \cite{Hong}, for example, 
the evaluation of the anomalous triangle diagram in the 
color-flavor-locked phase does not pick up the 
$N_c$ factor from the quark loop. 

Note that in Eqs.(\ref{L*}) 
$F/D=\langle n_0^2\rangle/\langle n_0^2+n_3^2 \rangle<1$ so that
$(e N_c)^*_{\gamma\pi^3}/(e N_c)_{\gamma\pi^3} 
< (e^2 N_c)^*_{\pi\gamma^2}/(e N_c)_{\pi\gamma^2}$, which 
implies that fewer colors would be effectively 
involved in the box diagram of process (c) than in the 
triangle diagram of process (b). 

\begin{figure}
\begin{center}
\includegraphics[width=5.4cm,height=8.1cm,angle=270]{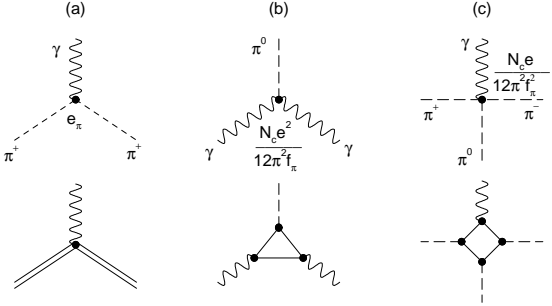}
\end{center}
\caption{The pion vertices for the processes and the corresponding 
quark processes.}
\end{figure}

Shown in Fig.3 is the density dependence of the parameters appearing in 
Eqs.(\ref{L*}). 
Quantities here are normalised with respect to their values in
free space and all 
are reduced by the effect of the baryonic medium. 
Only the pion mass appears stable at low density corresponding to 
the symmetry broken phase. The other quantities show quite 
strong dependence on the medium density even at low density.  
On the other hand, the pion mass drops quickly to zero in the pseudogap 
phase, while the other parameters exhibit plateau behaviour. 
In the chiral symmetry restored phase, all the parameters rapidly 
go to zero after $\rho_c$. 

\begin{figure}
\begin{center}
\includegraphics[width=5cm,height=7cm,angle=270]{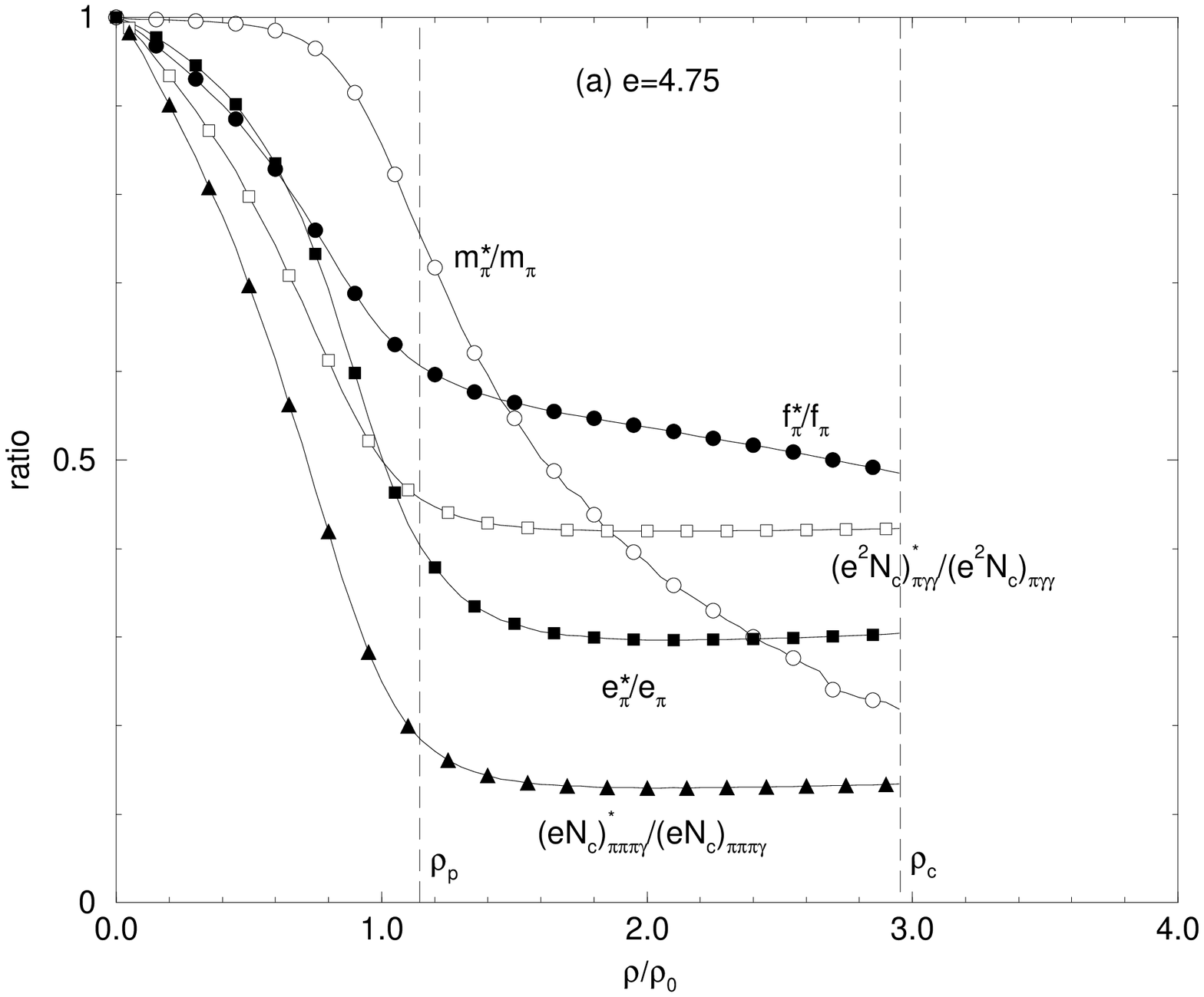}
\includegraphics[width=5cm,height=7cm,angle=270]{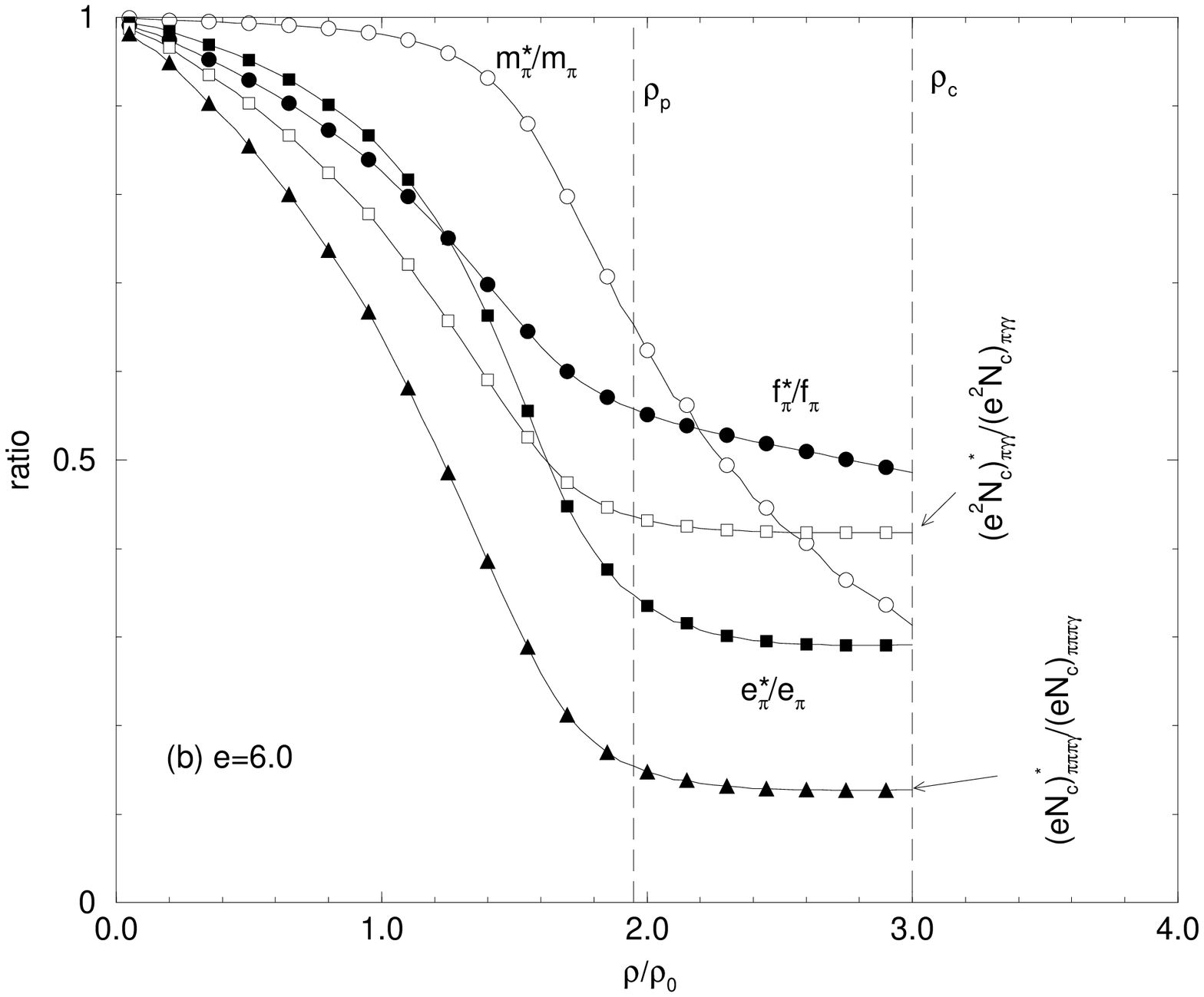}
\end{center}
\vskip -4ex
\caption{Ratios of the in-medium quantities to the free ones. 
In order to illustrate the strong parameter dependence 
of $\rho_p$, we present two results obtained with (a) $e=4.75$ 
and (b) $e=6.0$. The other parameters are fixed as $f_\pi$=93 MeV, 
$m_\pi$=138 MeV, $f_\chi$=240 MeV and $m_\chi$=720 MeV. }

\end{figure} 

As a consequence of changes in the in-medium parameters,  
the $\pi^0 \rightarrow \gamma\gamma$ decay width is modified,  
with the result  
\begin{equation}
\Gamma^*_{\pi^0\gamma\gamma} 
= \frac{m_{\pi}^{*3}}{64\pi} 
\left( \frac{(N_c e^2)^*_{\pi\gamma^2}}{12\pi^2 f^*_\pi}\right)^2.
\label{modifiedwidth}
\end{equation}
That this result is identical in structure
to Eq.(\ref{anomaly}) up to the replacements 
$m_{\pi}\rightarrow m_{\pi}^{*}, f_{\pi} \rightarrow \ f^{*}_{\pi}$ and
so on, is a consequence of the preservation of the chiral
anomaly in the presence of dense matter.  
Thus for example the pion mass coming from the phase factor 
of the process is replaced by the effective pion mass.
In obtaining Eq.(\ref{modifiedwidth}), we have used a very naive 
mean field approximation. 
The higher order contributions in the medium-generated 
potentials of Eqs.(\ref{potentials}) can be incorporated systematically.  
However, as shown in the previous works on the other properties of 
pion \cite{LPMRV03,LPRV04}, 
after the resummation they lead only a minor corrections compared to  
the zeroth order values that are nothing but those obtained 
in the  mean field approximation. 

Temperature effects could have been revealed, for example, 
in \cite{AEDGN94}  
from the pion loops, because the pion propagator at finite temperature 
differs from that at zero temperature by discrete Matsubara frequencies. 
One may extract a density dependence by evaluating similar pion loops 
by using the modified pion propagator with effective pion mass.   
We expect, however, that the result is less important than 
the dominant nonperturbative one through direct in-medium 
modification of the effective parameters.

Since the dilaton field is decoupled from the anomalous WZW term 
of the Lagrangian and is neutral, this particle cannot 
be involved directly in electromagnetic 
processes such as as $\pi^0 \rightarrow \gamma\gamma$. 
The presence of background medium could generate a 
$\pi-$matter$-\chi$ coupling, which can be 
absorbed into the resummation process like other terms 
higher order in the potentials Eqs.(\ref{potentials})
and leads to small corrections on top of the mean field
result we have pursued here.
However, as discussed in \cite{LPRV03}, 
the dilaton plays its most important role in the chiral 
symmetry restoration of the background matter.
Thus, the dilaton field is indirectly relevant to the process 
$\pi^0 \rightarrow \gamma\gamma$, as mentioned above,
by providing for a consistent context of chiral symmetry restoration
within which the anomalous process can be studied. We shall
return to this again below.

Numerical results for Eq.(\ref{modifiedwidth}) are presented in Fig.4.
One can see that the neutral pion decay process is strongly supressed 
as the matter density increases. This   
comes from the reduction in the pion mass (solid line
in the inset graph) and from the reduction in the strength of the 
corresponding vertex (dashed line), where the former plays more 
a dominant role in the pseudogap phase and the latter in the symmetry 
broken phase.

We now compare our results to other work specifically focussing
on qualitative behaviour, whether the decay width is enhanced
or suppressed. 
To this end, it is useful to state the three broad
approaches within which temperature/density dependence
is reflected in pion properties and chiral symmetry restoration
at the critical point. Phenomenologically these different
approaches centre around different ways of realising
that the ratio of $m_{\pi} f_{\pi}$ in medium vs. in vacuo
must be equal to the corresponding ratio for $m_q \langle {\bar q} q\rangle$
(Gell-Mann--Oakes--Renner relation), 
which must in turn vanish if chiral symmetry is restored
at the phase transition. 
Also relevant is that upon restoration of the symmetry, the
pion and its chiral partner the $\sigma$, here represented
by the dilaton, must become degenerate.
Thus we can label as approach A the case of the pion mass increasing to become
degenerate with its chiral partner, while $f_{\pi}$ to
approaches zero to fulfill the ratio of GMOR relations.
Approach B would have both the pion and its chiral
partner decreasing in mass to become eventually degenerate,
precisely the case for our work with the pion and dilaton
becoming degenerate. No experimental evidence yet invalidates
one of these against the other.   
In comparing our results with the other aproaches mentioned
in the introduction the difference in the final result
for the decay width will ultimately depend on which approach
(A or B) has been realised for chiral symmetry restoration.
Furthermore, to compare with works that only
include finite temperature effects we will follow the general consensus
and take our density dependence as qualitatively indicative also
of the behaviour as a function of temperature.

Firstly, our decay width suppression disagrees with \cite{GNAE93}
which however does not incorporate medium (temperature
or density) dependences in quantities such as
the pion mass and the decay constant. To that extent
\cite{GNAE93} corresponds neither to approach A or B above
and in fact can only be relevant for low temperature or
density far away from any transition to chiral symmetry restoration.  
Our result agrees with \cite{CRK03} in the decay width, though in our
calculation the pion mass decreases with density while 
\cite{CRK03} have chiral symmetry restoration realised via
approach A: the pion mass increases to that of the $\sigma$
meson to achieve symmetry restoration.  
Our approach is closest to \cite{Caldas} in that
the temperature/density dependence of the pion mass and the decay 
width are incorporated. In particular a vanishing pion mass at the 
phase transition point corresponds to case II of \cite{Caldas}
and approach B delineated above.   
However, in \cite{Caldas} the mass is almost constant 
in temperature and density until the phase transition and then 
either suddenly increasing (Case I) or decreasing (II). Consequently, the 
temperature/density dependence of the pion decay constant 
plays the most important role, which makes the decay width 
enhanced over some range in both cases but near the phase transition dropping 
suddenly to zero in case II.

In our approach the pion mass is almost stable in the chiral 
symmetry broken phase, then decreases steadily through 
the pseudogap phase to vanish at the transition to
the symmetry restored phase, thereby realising
restoration via approach B, as above. 
On the other hand, both the effective pion decay 
constant $f_\pi^*$ and the effective values for $(N_c e^2)^*$  
decrease monotonically in the chiral symmetry broken phase but 
become stable in the pseudogap phase. If we do not take into 
account the change in $(N_c e^2)$, the decreasing pion decay 
constant alone would lead to an enhancement of the decay width 
(as in case II of \cite{Caldas}) in the low density regime 
where the chiral symmetry is still broken. 
Since $(N_c e^2)^2$ decreases faster than $f_\pi^*$, 
the decay width becomes decreasing in that region. 
This in medium modification of
 $(N_c e^2)$ then is the key result distinguishing our
result for the $\pi^0\rightarrow \gamma \gamma$ decay width 
from others while realising chiral symmetry
restoration via decreasing pion mass. 
Fig.3 and 4 illustrate the degree of sensitivity
of our results on the Skyrme parameters. 
Either way, we see in the two curves of Fig.4 that qualitatively  
the decay width suppression is robust against such variations. 
This result smoothly matches to the results of \cite{CFL}
where the analogous process of generalised pion decay
$\tilde{\pi}^0 \rightarrow \tilde{\gamma}\tilde{\gamma}$
is also suppressed in the color-flavor-locking phase 
which would take place at higher density above $\rho_c$.

\begin{figure}
\begin{center}
\includegraphics[width=5cm,height=7cm,angle=270]{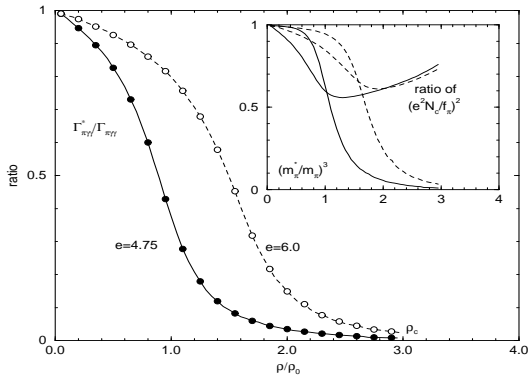}
\end{center}
\vskip -4ex
\caption{Ratios of the decay width of $\pi^0 \rightarrow\gamma\gamma$ to 
the free space value. The curves of the inset graph are the ratios 
$(m^*_\pi/m_\pi)^3$ and $((eN_c/f_\pi)^*/(eN_c/f_\pi))^2$.
Again the results obtained with two different values of 
the Skyrme parameter: $e$=4.75(solid lines) and $e$=6.0 (dashed lines)}
\end{figure} 

\section{Conclusions}
The main result of this work is given in Fig.4 which shows
suppression of the process $\pi^0\rightarrow \gamma \gamma$
computed in a unified framework of the Skyrme model
for pion fluctuations in a dense baryonic matter background.
The in-medium dependences of the pion mass and decay
constant arise in the same effective chiral theory as the anomaly 
which drives this process. The key new result emerging
from this Skyrme approach is that  
$(N_c e^2)^*$ can also be medium dependent in the {\it hadronic
phase} just as it does in the CFL phase \cite{CFL}. 

We repeat that this study of $\pi^0\rightarrow \gamma \gamma$
decay at finite density using skyrmions
is qualitative, due to the naivety in the crystal structure
of the background baryons, the absence of a treatment of the Fermi
statistics of the nuclear matter and the absence of loop
contributions.

Our result would appear to be not so relevant to 
relativistic heavy ion collisions, for example the
observation of suppression of neutral pions created in central
vs. peripheral collisions
as seen at PHENIX \cite{PHENIX}.
However, the scenario of low temperature and high density
matter is particularly relevant to the context of
compact stars. Applications of these results to
this problem are underway.

\section*{Acknowledgements}
B-Y.P. is grateful for the hospitality of the Special Research Centre for
the Subatomic Structure of Matter at the University of Adelaide,
where part of this work has been done. A.C.K. is supported by 
the Australian Research Council. The authors are
grateful to the referee for constructive comments
and additional references and to Prof. B. Hong for discussions 
on heavy ion collisions.


\begin{thebibliography}{11}
\bibitem{LPMRV03}
H.-J. Lee, B.-Y. Park, D.-P. Min, M. Rho and V. Vento, 
   Nucl.Phys. {\bf A723} (2003) 427.
\bibitem{LPRV03}
H.-J. Lee, B.-Y. Park, M. Rho and V. Vento, 
   Nucl. Phys. {\bf A726} (2003) 69.
\bibitem{LPRV04}
H.-J. Lee, B.-Y. Park, M. Rho and V. Vento, 
   Nucl. Phys. {\bf A741} (2004) 161.
\bibitem{PRV04}
B.-Y. Park, M. Rho and V. Vento,
   Nucl. Phys. {\bf A 736} (2004) 129.

\bibitem{SkyrmionCrystal}
I. Klebanov, 
   Nucl. Phys. {\bf B262} (1985) 133;
L. Castillejo, P.S.J. Jones, A. D. Jackson, 
       J.J.M. Verbaarschot and A. Jackson,
   Nucl. Phys. {\bf A501} (1989) 801;
M. Kugler and S. Shtrikman,
   Phys. Lett. {\bf B208} (1988) 491;
   Phys. Rev. {\bf D40} (1989) 3421.
A.D. Jackson, A. Wirzba, and L. Castillejo,
   Nucl. Phys. {\bf A486} (1988) 634; 
   Phys. Lett. {\bf B198} (1987) 315;
H. Forkel, A.D. Jackson, M. Rho, and C. Weiss,
   Nucl. Phys. {\bf A504} (1989) 818;
D.I. Diakonov and A.D. Mirlin, 
   Sov. J. Nucl. Phys. {\bf 47} (1988) 424.

\bibitem{Astrophysics}
For a review, see
M. Prakash, J. M. Lattimer, R. F. Sawyer, and R. R. Volkas,
   Ann. Rev. Nucl. Part. Sci. {\bf 51} (2001) 295.

\bibitem{ANV03}
F. Arretche, A. A. Natale and D. N. Voskresensky, 
   Phys. Rev. C {\bf 68}, (2003) 035807.

\bibitem{FiniteT}
H. Itoyama and A. H. Muller, 
   Nucl. Phys. {\bf B218} (1983) 349;
T. Hashimoto, K. Hirose, T. Kanki and O. Miyamura, 
   Phys. Rev. {\bf D37} (1988) 3331;
C.Contreras, M.Loewe, 
   Z.\ Phys. {\bf C40} (1988) 253;
A.~A.~Rawlinson, D.~Jackson and R.~J.~Crewther,
   Z.\ Phys. {\bf C56} (1992) 679;
Bi Ping-Zhen and J. Rafelski, 
   Mod. Phys. Lett. {\bf 85} (2000) 3595.

\bibitem{Pi96}
R. D. Pisarski, 
   Phys.Rev.Lett. 76 (1996) 3084.

\bibitem{PTT97}
R. D. Pisarski, T.L. Trueman, M.H.G. Tytgat, 
   Phys. Rev. {\bf D56} (1997) 7077.

\bibitem{AEDGN94} 
R. F. Alvarez-Estrada, A.Dobado, A.G\'omez Nicola,
   Phys. Lett. {\bf B324} (1994) 345.

\bibitem{GNAE93} 
A. G\'omez Nicola, R.F.Alvarez-Estrada, 
   Z. Phys. {\bf C60} (1993) 711.

\bibitem{CRK03} 
P. Costa, M. C. Ruivo, Yu. L. Kalinovsky,
   Phys. Lett. {\bf B577} (2003) 129; {\it Erratum-ibid} {\bf B581}
(2004) 274. 

\bibitem{Caldas} 
   H. Caldas, Phys. Rev. {\bf C69} (2004) 035204.

\bibitem{CFL}
M. A. Nowak, M. Rho, A. Wirzba and I. Zahed,
   Phys. Lett. {\bf B497} (2001) 85.

\bibitem{Hong}
D. K. Hong, M. Rho and I. Zahed, 
   Phys. Lett. {\bf B468} (1999) 111.

\bibitem{ScaledLagrangian}
J. Ellis and J. Lanik,
   Phys. Lett. {\bf B150} (1985) 289;
V.A. Novikov, M.A. Shifman, A.I. Vainshtein, V.I. Zakharov,
   Nucl. Phys. {\bf B165} (1980) 67;
H. Gomm, P. Jain, R. Johnson, and J. Schechter,
   Phys. Rev. {\bf D33} (1986) 3476;
P. Jain, R. Johnson and J. Schechter,
   Phys. Rev. {\bf D35} (1987) 2230;
G.E. Brown and M. Rho,
   Phys. Rev. Lett. {\bf 66} (1991) 2720. 

\bibitem{Witten83}
E. Witten, 
   Nucl. Phys. {\bf B223} (1983) 422.

\bibitem{Vento03}
See, for example, V. Vento, hep-ph/0401218. 

\bibitem{dilaton}
R. J. Furnstahl, H.-B. Tang and S. D. Serot,
  Phys. Rev. {\bf C52} (1995) 1368;
S.R. Beane and U. van Kolck, 
  Phys. Lett. {\bf B328} (1994) 137;
C. Song, G. E. Brown, D.-P. Min and M. Rho,
  Phys. Rev. {\bf C56} (1997) 2244.

\bibitem{JM83}
A. D. Jackson and M. Rho,
  Phys. Rev. Lett. {\bf 51} (1983) 751.

\bibitem{ANW83}
G. Adkins, C. Nappi, E. Witten,
  Nucl. Phys. {\bf B228} (1983) 552.

\bibitem{Walet02}
  Oliver Schwindt, Niels R. Walet, 
    ``Soliton Systems at Finite Temperatures and Finite Densities", 
    hep-ph/0201203

\bibitem{PHENIX}
K. Adcox {\it et al\/}, PHENIX Collaboration,
   Phys. Rev. Lett. {\bf 88} (2002) 022301;
   {\it ibid}, 242301.
\end{thebibliography}
\end{document}